# Accurate Prediction of Bonding Properties by A Machine Learning–based Model using Isolated States Before Bonding

Eiki Suzuki, Kiyou Shibata, and Teruyasu Mizoguchi*

*Institute of Industrial Science, The University of Tokyo, 153-8505 Tokyo, Japan*

**\*Corresponding author.** Email: teru@iis.u-tokyo.ac.jp

**Abstract**

Given the strong dependence of material structure and properties on the length and strength of constituent bonds and the fact that surface adsorption and chemical reactions are initiated by the formation of bonds between two systems, bonding parameters are of key importance for material design and industrial processes. In this study, a machine learning (ML)-based model is used to accurately predict bonding properties from information pertaining to isolated systems before bonding. This model employs the density of states (DOS) before bond formation as the ML descriptor and accurately predicts binding energy, bond distance, covalent electron amount, and Fermi energy even when only 20% of the whole dataset is used for training. The results show that the DOS of isolated systems before bonding is a powerful descriptor for the accurate prediction of bonding and adsorption properties.

TOC

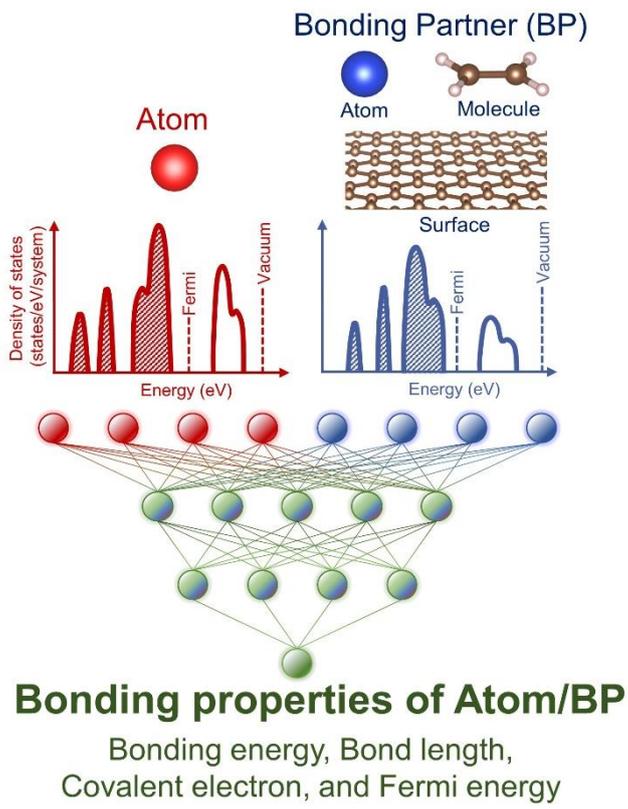
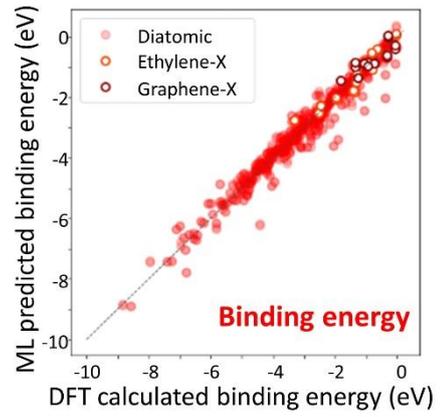
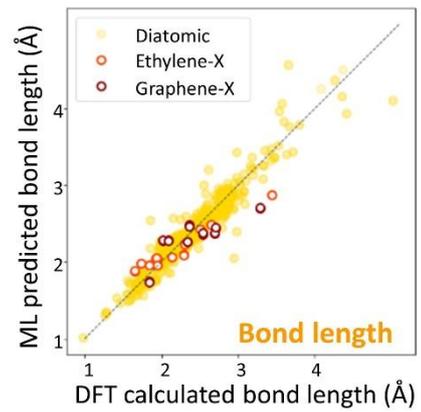

The molecular and crystal structures and electronic structure of materials are largely influenced by characters of bonding between their constituent atoms. Bond length and strength and their formations play crucial roles in surface adsorption processes and chemical (e.g., catalytic) activities.[1–5] Although numerous first-principles electronic structure simulations and spectroscopic experiments have been conducted to probe bonding character, there is still much to be learned in this regard.[1,3,6–9]

To obtain a working understanding of bonding character, Linus Pauling and others introduced electronegativity, a parameter that describes the strength and nature of bonds between atoms,[10–13] but, when used on its own, is poorly suited for the accurate quantitative description of bonding between two chemical species.

If the parameters of bonded systems can be predicted from those of the corresponding isolated systems before bonding, some combinations may enable generalization and thus allow one to make predictions for a vast array of other bonded systems comprising combinations of isolated systems. This advantage cannot be obtained using the parameters of bonded systems as explanatory variables. Using isolated systems also effective from the perspective of efficiency since obtaining the parameters of bonded systems requires much higher computational cost than isolated system due to structural optimization. Furthermore, such accurate predictions should facilitate the selection of materials and the design of their functions related to adsorption and chemical reactions.

In this study, inspired by the recent use of machine learning (ML) to extract useful information from a large amount of electronic structure data,[14–19] we developed a new ML-based method of bonding property prediction. The constructed model quantitatively predicted bonding properties, such as bond length and strength, from the electronic densities of states (DOSs) of isolated systems before bonding.

The DOS is an energy distribution of the states that can be occupied by electrons and is usually used to describe the electronic structure of materials. Previously, the DOS has been used as a ML descriptor to predict the DOS at the Fermi level and other DOS features.[20,21] Furthermore, the DOS was recently used for prediction of the binding energy of molecules on metal surface[22,23]. However, no attempt has been made to predict several bonding properties, including binding energy, bond length, and chemical bonding, from the DOSs of isolated systems before bonding.

Figure 1 schematically shows the concept of this study. The DOSs of isolated systems before bonding were used as inputs to predict four bonding properties, viz. binding energy, bond length, covalent electron amount, and Fermi energy after bonding. To understand bonding between atoms in different bonding pairs, we created databases for three types of systems, demonstrating that the DOSs of isolated systems before bonding are powerful descriptors for predicting the properties of the above bonding systems.

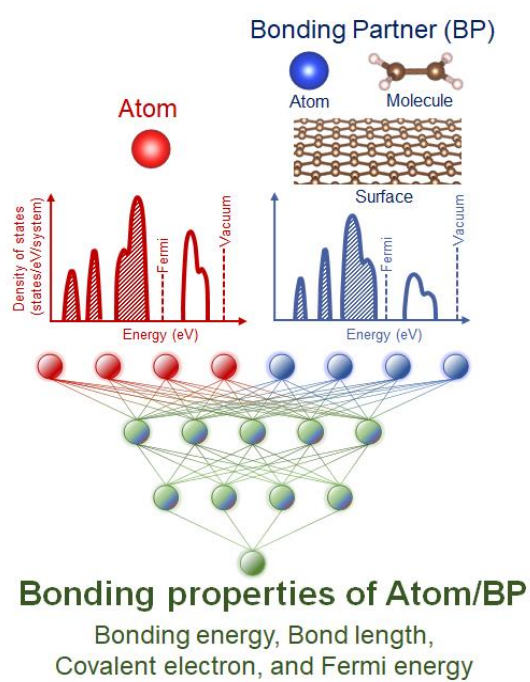

Fig.1. Schematic illustration of the present study. The ML predictions for various bonding properties between atom and bonding partner (BP) based on density of states (DOSs) of isolated systems before bond creation.

First-principles calculations were performed based on density functional theory (DFT) using the projector-augmented wave (PAW) method with a plane-wave basis set, as implemented in the Vienna *ab initio* Simulation Package (VASP).[24] Semi-core orbitals were included in valence. Spin-polarized calculations were performed, but the spin–orbit interaction was not considered. The PAW potentials incorporating scaler relativistic (mass velocity and Darwin terms) were used.

Effects of exchange correlation and van der Waals (vdW) force functionals on the simulation of the surface adsorption have been extensively investigated. [6–8,25] In this study, as isolated atoms, molecule, bulk, and adsorption systems are needed to be simulated, the density functional was selected by following criteria: 1) The calculation of some isolated systems are stably converged and the calculated total energies are negative (stable). 2) The accuracy of the functionals was estimated using a simulated interlayer distance of bulk graphite. We have examined six functionals, including Perdew-Burke-Ernzerhof (PBE)[26], SCAN+rVV10[25], Tkatchenko-Scheffler[27], DFT-D2[28], optPBE[29], and optB88[30]. Only SCAN+rVV10 and PBE show stable convergence with negative energies for all isolated systems, whereas other functionals show positive energies or un-converged results in some systems. The errors in simulating the interlayer distance of graphite were +1.8% and +19.5% for SCAN-rVV10 and PBE, respectively. Based on this consideration, SCAN+rVV10 was used as the vdW density functional.

To obtain basic and important knowledges on bond formations and adsorption, we selected following three types of model systems, viz. 1) diatomic systems consisting of various elements, 2) an atom adsorbed on an ethylene molecule, and 3) an atom adsorbed on a bulk surface modeled by a graphene sheet. Ethylene and graphene were selected because of their technological importance, as exemplified by the production of polyethylene[31] and next-generation electric devices[32]. Simple adsorption positions were employed to systematically compare adsorption behaviors. For 2), we investigated the adsorption of an atom from the normal direction to the center of the C=C bond of a planar ethylene molecule. For 3), we considered atom adsorption at an on-top site, i.e., above a carbon atom. Notably, the most stable atomic sites varied from system to system. Furthermore, to keep the selected adsorption systems, some atoms were constrained in cases 2) and 3) during their atomic optimizations. The structure of ethylene was fixed to avoid double bond breakage, and some graphene atoms were fixed to preserve adsorption at the on-top site. In addition to these atom/bonding partner (BP) systems, isolated systems, viz. elements from H ($Z = 1$) to Rn ($Z = 86$), an isolated ethylene molecule, and graphene, were separately modeled. Each model was positioned around the supercell center.

To minimize interactions between mirror atoms, all calculations were performed using sufficiently large supercells, e.g., 15 Å × 15 Å × 15 Å cubic cells for diatomic and molecular adsorption systems and 7 Å × 7 Å × 14.7 Å supercells having a graphene sheet composed of 18 atoms for graphene adsorption systems. An energy cutoff of 500 eV and a 4 × 4 × 1 Γ-centered $k$-point grid were used for all calculations. The atomic structures were relaxed until their residual forces became less than 0.05 eV/A.

Binding energy, Fermi energy, covalent electron amount, bond distance, and total DOS were obtained for all systems except for those without bonding and those with non-converging calculations (possibly due to unstable interactions).

The binding energy of diatomic/adsorption systems ($E_{binding}$) was defined as

$$E_{binding} = E_{whole} - (E_{Atom} + E_{BP}),$$

where $E_{whole}$ is the total energy of the optimized system containing the adsorbing/bonding atom and the BP, $E_{Atom}$ is the total energy of the bonding/adsorbing atom, and $E_{BP}$ is the total energy of the isolated atom (diatomic system), an isolated ethylene molecule, or an isolated graphene sheet.

To validate our simulations, we compared the calculated binding energies and bond lengths with those reported previously. The bond length and biding energy of diatomic molecules were compared with database and confirmed that the present simulation well reproduced the latter values as shown in Fig.S1[33,34]. The atom adsorption on graphene has also been reported[35–37], and our simulation reproduces the results obtained by the similar calculation conditions[37].

The amount of covalent electrons ($Q_C$) observed upon going from isolated systems to diatomic/adsorption systems are defined as

$$Q_C = \{(Q_{Atom'} + Q_{BP'}) - (Q_{Atom} + Q_{BP})\},$$

where $Q_{Atom}$ and $Q_{BP}$ are the site-projected charges of the isolated atom and the BP (atom, ethylene, or graphene), respectively, while $Q_{Atom'}$ and $Q_{BP'}$ are the site-projected charges of the adsorbed atom and the BP (atom, ethylene, or graphene) of diatomic/adsorption systems, respectively.[38]

The calculated total DOSs of atom/BP and isolated systems were extracted from −300 to 0 eV (vacuum level) and transformed into vectors with 1500 dimensions by linear interpolation. Gaussian smoothing with 1 eV was applied for all DOSs. A feed-forward neural network comprising two concatenation layers and three fully connected layers was adopted as the ML model for predictions. The first layer concatenated the up- and down-spin DOSs of each atom, while the third layer concatenated the two branches from the DOSs of isolated systems. The second and fourth layers were fully connected layers with 64 output neurons and featured a sigmoid activation function. The fifth layer is directly connected to the output using a linear activation function in all coupled layers. The model hyperparameters, number of nodes, number of layers, and activation functions were optimized using Bayesian optimization. The TensorFlow framework, which is a high-level neural network API,[39] was used in our neural network model. The model adopted the mean-squared error as the loss function and was trained using the *Adam* stochastic optimization method.[40]

The up- and down-spin DOSs of two isolated atoms and BPs before bonding were concatenated and used as input data. For input into the model, each DOS was scaled such that its integrated value up to Fermi energy was proportional to the number of its valence electrons considered in DFT calculations. The respective standardized values of binding energy, bond length, covalent electron amount, and Fermi energy were used as output data. As our model was not symmetric about the two input DOSs, the dataset was chosen irrespective of the order of atoms, i.e., including both atom/BP and BP/atom combinations. Excluding systems without bonding and non-converging systems, the total number of atom/atom, atom/ethylene, and atom/graphene combinations equaled 1973, 56, and 47, respectively. By adding atom/BP and BP/atom combinations (except for homonuclear diatomic molecules), we obtained a dataset containing 4093 systems.

To validate the prediction model, we have performed 5fold cross-validation, in which four-fifths of the training set was used to train the models, and one-fifth was used as validation data. The average of the five output predictions was used to determine the final prediction accuracy. The objective variables of the training set were standardized by their mean and standard deviation.

The dataset was split into training and test sets, with the role of the training/test data ratio discussed later. The test set must not include the combinations appearing in the training set, i.e., the equivalent combinations of both atom/BP and BP/atom were located in either one or the other set. For this data splitting, we first classify the atom/BP pairs to be the training and test and added corresponding BP/atom pairs to the respective categories.

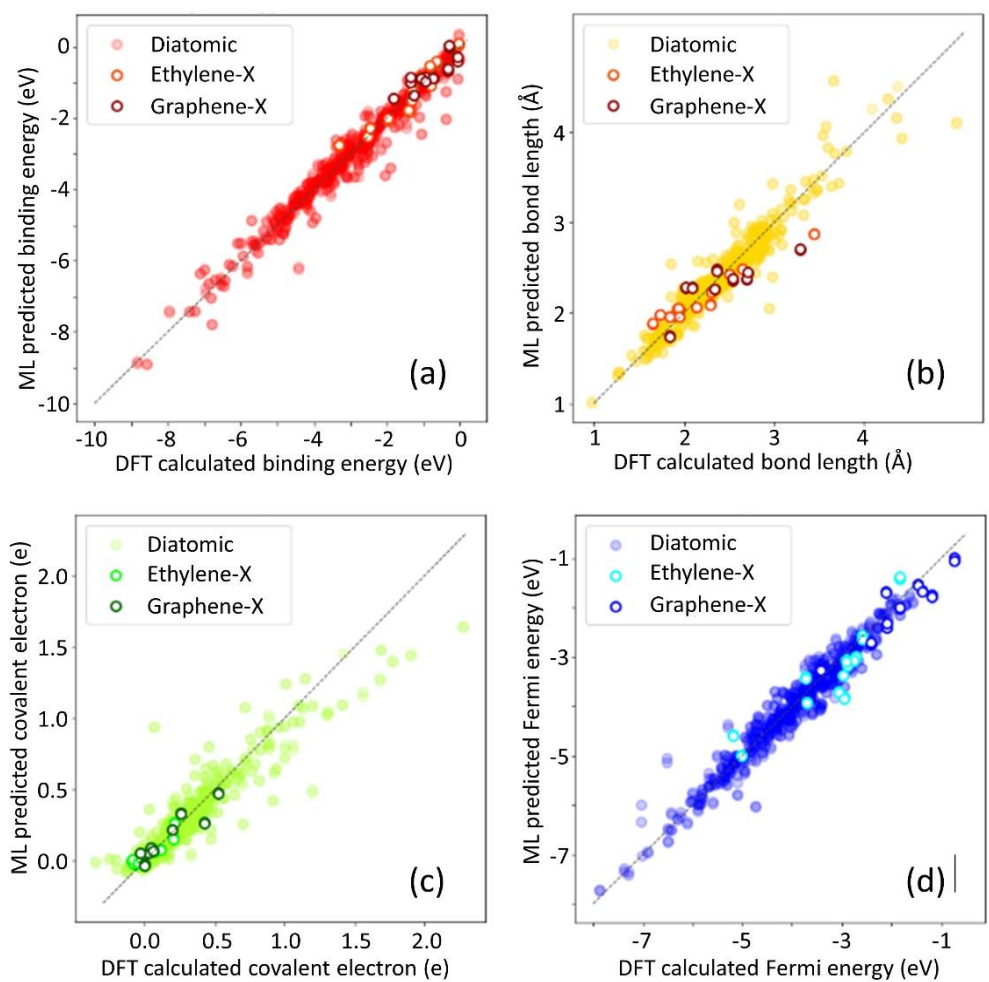

**Fig. 2.** Parity plot showing the distribution of predicted vs. calculated values of (a) binding energy, (b) bond distance, (c) covalent electron amount, and (d)

Fermi energy for the unobserved species of stable diatomic molecules. The test data account for 20% of the whole data set.

Figure 2(a) shows the parity plots of binding energy predictions for diatomic molecules, atom/ethylene adsorption (ethylene-X), and atom/graphene adsorption (graphene-X). The test data accounted for 20% of the whole data set and, as mentioned above, did not include the combinations appearing in the training data set. The validation plot featured $R^2$ = 0.963 and a mean absolute error (MAE) of 0.224 eV, indicating that all binding energies, irrespective of the system, were accurately predicted using the DOSs of isolated systems as descriptors. Although no constraint was used, the MAE for atom/BP and BP/atom bonding energies was approximately 0.08 eV, indicating that the constructed prediction model correctly grasps the kind of elements and BP from their input DOSs.

The data distributions obtained for diatomic molecules, ethylene-X, and graphene-X cases were different. The binding energies for adsorption systems (ethylene-X and graphene-X) were relatively low (close to zero), whereas a broad range of values was observed for diatomic systems, -9~0eV. As graphene and ethylene have similar carbon-carbon π-bonding characters, their isolated atom binding energies can be similar to each other. The similar prediction of the binding energy using DOS has been recently reported, but we have applied this method to predict other adsorption properties, such as bond length, Fermi energy, and covalent electron amount as below [22,23].

In addition to binding energy, the stable bond lengths between atoms and BPs were also correctly predicted (Fig. 2(b)). The corresponding $R^2$ = 0.834 (MAE = 0.105 Å) was very high, whereas the data distributions were similar to each other irrespective of the atom/BP system, unlike in the case of binding energy. This difference in data distribution may indicate that bond length is not much sensitive to bonding/adsorption systems, but is mainly determined by the combination of atomic species. On the other hand, binding energy is correlated to the state of the system, e.g., isolated atom, molecule, or solid.

Similar to the cases of binding energy and bond length, the amount of covalent electrons (Fig. 2(c)) and Fermi energy (Fig. 2(d)) were predicted with high accuracies, with the respective $R^2$ (MAE) values equaling 0.866 (0.075 electron) and 0.945 (0.190 eV). Furthermore, the data distribution for bonding/adsorption systems depended on their properties.

These results commonly indicate that the DOSs of isolated systems before bonding is the powerful descriptors for predicting bonding/adsorption properties, including binding energy, stable bond length, amount of covalent electrons, and Fermi energy.

Here, the physics behind the descriptor proposed in this study, the DOSs of isolated system, is discussed. These bonding properties are related to orbital–orbital interactions. The basic molecular orbital theory suggests that the magnitude of these interactions is mainly determined by the difference between orbital energy levels, with orbital overlap depending on orbital type. These two pieces of information can be determined from DOS shape, i.e. DOS position and intensity represent the orbital energy level and type, respectively. Thus, DOSs can be used as descriptors to predict these bonding/adsorption properties.

On the other hand, the relationships between Fermi energy and DOS shape have not been explicitly mentioned. However, our results suggest that some buried information on Fermi energy is present in the DOS profile, which can be an indication of relationship between Fermi energy and the DOS profile through orbital energy levels and number of electrons.

It is worth noting that our model could predict the properties of stable systems. For two isolated systems separated by a known distance, one can quantitatively calculate the binding state using approaches such as the molecular orbital method. When only the two isolated systems are known, however, even state-of-the-art first-principles techniques require iterative calculations to obtain stable system parameters, which suggests that big data and ML approaches can be effectively used for previously not considered tasks.

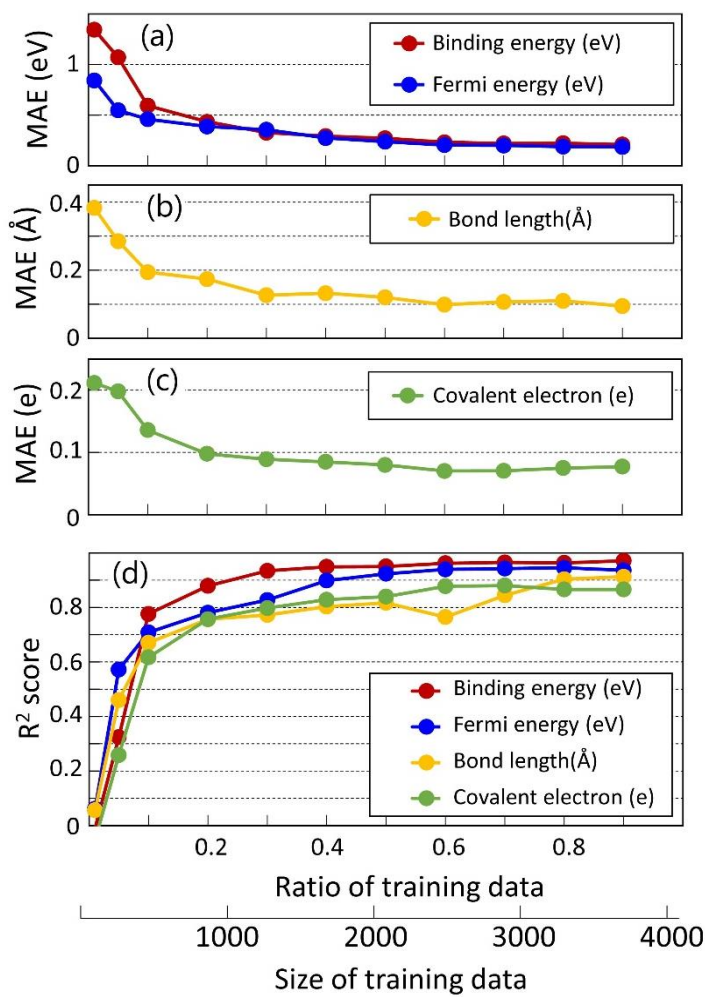

**Fig. 3.** Correlation between the training set fraction and prediction accuracies. The accuracy was estimated by mean absolute errors (MAE) and $R^2$ for selected properties. (a) MAE for binding energy, Fermi energy, (b) bond distance, and (c)

covalent electron. (d) $R^2$ for all properties.

Finally, we investigated prediction accuracy, $R^2$ and MAE, as a function of the training set fraction (MAE for binding energy and Fermi energy in Fig.3(a), bond length in Fig.3(b), covalent electron amount in Fig.3(c), and $R^2$ in Fig. 3(d)), revealing that these two parameters are positively correlated, with $R^2$ exceeding 0.7 when the training set fraction exceeds 0.2.

Even models with a training set fraction of 0.2 yielded $R^2 > 0.7$ for binding energy (MAE=0.44eV), bond length (MAE=0.17Å), amount of covalent electrons (MAE=0.10e), and Fermi energy (MAE=0.39eV), which suggested the high generalization capability of our model and implied that it can be applied to other systems obtained by combination of additional isolated systems by means of transfer learning. Thus, our study revealed that the DOSs of isolated systems contain abundant information and can be powerful descriptors for predicting bonding/adsorption properties.

In summary, a ML approach was used to predict the chemical bonding, binding energy, bond distance, covalent electron amount, and Fermi energy of bonding/adsorption systems from the DOSs of isolated systems before bond creation. The constructed model achieved very accurate predictions of all properties even when the training data amounted to 20% of the whole data set.

The pairs of DOSs of isolated systems before bond creation were concluded to contain much useful information on the bonded systems (binding energy, bond length, covalent electron, and Fermi energy) and be powerful descriptors for predicting numerous bonding properties. As the developed approach has a high generalization capability, it can be applied to the bonding of larger molecules such as nano clusters, nanowires, and catalytic molecules, and the adsorption of atoms at bulk surfaces.

## Author Contributions

T.M. and K.S. supervised this research and directed the calculations. E.K. and K.S. performed simulations under supervision of T.M. All authors contributed to the scientific discussion and commented on the results. The manuscript is written by all authors.

## Competing Interests Statement

The authors declare no competing financial interests.

## Acknowledgements

This study was supported by the Ministry of Education, Culture, Sports, Science and Technology (MEXT; Grant Nos. 17H06094, 19H00818, and 19H05787), and a special grant of the Institute of Industrial Science, The University of Tokyo (Tenkai5504850104).

**Figure captions**

**Fig. 1.** Fig.1. Schematic illustration of the present study. The ML predictions for various bonding properties between atom and bonding partner (BP) based on density of states (DOSs) of isolated systems before bond creation.

**Fig. 2.** Parity plot showing the distribution of predicted vs. calculated values of (a) binding energy, (b) bond distance, (c) covalent electron amount, and (d) Fermi energy for the unobserved species of stable diatomic molecules. The test data account for 20% of the whole data set.

**Fig. 3.** Correlation between the training set fraction and prediction accuracies. The accuracy was estimated by mean absolute errors (MAE) and $R^2$ for selected properties. (a) MAE for binding energy, Fermi energy, (b) bond distance, and (c) covalent electron. (d) $R^2$ for all properties.